\documentclass[aps,prl,twocolumn,showpacs]{revtex4}
\usepackage{graphicx}
\usepackage{bm}

\begin{document}

%\preprint{\sf Version 1 (\today)}
\title{Theory of Four-dimensional Fractional Quantum Hall States}
\author{Chyh-Hong Chern}
\email{chern@appi.t.u-tokyo.ac.jp}

\affiliation{ERATO-SSS, Department of Applied Physics, University
of Tokyo, Bunkyo-ku, Tokyo 113-8656, Japan}

\begin{abstract}
We propose a pseudo-potential Hamiltonian for the Zhang-Hu's
generalized fractional quantum Hall states to be the exact and
unique ground states. Analogously to Laughlin's quasi-hole
(quasi-particle), the excitations in the generalized fractional
quantum Hall states are extended objects. They are vortex-like
excitations with fractional charges $+(-)1/m^3$ in the total
configuration space CP$^3$. The density correlation function of
the Zhang-Hu states indicates that they are incompressible liquid.
\end{abstract}

%\pacs{Valid PACS appear here}
\maketitle

Recently, the four-dimensional generalization of the quantum Hall
effect (4DQHE) proposed by Zhang and Hu (ZH) has drawn
considerable attentions in both condensed matter physics and high
energy physics communities \cite{Zhang2001Science}.  The projected
generalized Hall conductivity is considered to be the root of the
spin Hall effect in the $p$-type semiconductor
\cite{Murakami2003Science}.  When applying the electric field to
the $p$-type semiconductor, the dissipationless spin current flows
perpendicular to the direction of the electric field.  The purely
electrical manipulation of the spin current is one of the ultimate
goals for the applications in the information technology. On the
other hand, 4DQHE intrigues people by its topological
property\cite{bernevig-2002-300} and the deep relation with the
division algebra\cite{bernevig-2003-91}.  Furthermore, numerous
works have been done in the language of the modern string theory
\cite{Fabinger2002JHEP, Henriette2002}, part of which are
motivated by the massless edge excitations which carry integer
spins with helicity\cite{Hu2002PRB}. The underlying
non-commutative feature of ZH theory also fosters intensive
studies in the non-commutative geometry in the higher dimensional
compact spaces \cite{Kimura2002NPB}.

Besides the generalized \emph{integer} quantum Hall effect, ZH
also proposed a wave function with fractional filling, which they
referred to the generalized \emph{fractional} quantum Hall states
(FQHS). In two dimensions, the integer QHE can be well understood
by the one-body quantum mechanics but in the fractional case it is
the two-body repulsive Coulombic interaction which plays the
dominant role for the new states of matter possessing the
excitations with fractional charges and fractional
statistics\cite{Laughlin1983PRL}.  Similarly, the ZH fractional
quantum Hall states are believed to have many interesting
properties, namely being disordered in higher dimensions and the
excitations with fractional charges.  Besides, the edge modes of
the FQHS are still far from clear.  Therefore, it is urgent to
understand the many-body interaction in the ZH FQHS. In this
paper, we will try to answer this question. Namely, we propose the
two-body interaction Hamiltonian for the FQHS to be the
non-degenerate ground state. Furthermore, the correspondent
Laughlin's quasi-particle (quasi-hole) in the 4D FQHS can be shown
as a four-dimensional extended object which can be considered as
the vortex in the total configuration space CP$^3$.  In addition,
we will show that the vortex-like excitations carry the fractional
charge $+(-) 1/m^3$.

Let us start with a brief review of ZH's construction.  The
non-consecutive jump from two dimensions to four dimensions
results from the underlying algebraic structures. In two
dimensions, the two-dimensional complex spinor coordinate
$\phi^\alpha$ used to construct the coherent state on the
two-sphere can be introduced by the first Hopf map, that is $X_i/R
= \bar{\phi}^{\alpha}(\sigma_i)_{\alpha\beta}\phi^\beta$, where
$X_i$ are the coordinates on the two-sphere, $R$ is the radius,
and $\sigma_i$ are the Pauli matrices.  ZH generalized it by
considering the second Hopf map, which is $X_a/R=\bar{\psi}^\alpha
(\Gamma_a)_{\alpha\beta}\psi^\beta$, where $X_a$ are the
coordinates on the four-sphere, $R$ is the radius, and $\Gamma_a$
are the SO(5) Gamma matrices given by
\begin{eqnarray}
\Gamma^{i}=\left(\begin{array}{cc} 0 & i\sigma_i \\ -i\sigma_i & 0
\end{array}\right),\ \Gamma^4=\left(\begin{array}{cc} 0 & 1 \\ 1 & 0
\end{array}\right), \ \Gamma^5=\left(\begin{array}{cc} 1 & 0 \\ 0 & -1
\end{array}\right)
\end{eqnarray}
where $i$ is from 1 to 3.  An explicit solution $\psi^\alpha$ of
the second Hopf map can be obtained as
\begin{eqnarray}
&&\left(\begin{array}{c} \psi^1 \\ \psi^2
\end{array}\right)=\sqrt{\frac{R+X_5}{2R}}\left(\begin{array}{c}\phi^1\\\phi^2\end{array}\right),\nonumber \\&&\left(\begin{array}{c} \psi^3 \\
\psi^4
\end{array}\right)=\sqrt{\frac{1}{2R(R+X_5)}}(X_4-iX_i\sigma_i)\left(\begin{array}{c}\phi^1\\\phi^2\end{array}\right)
\label{Eq:so5_fundamental_spinor}
\end{eqnarray}
where implicit summation is assumed and ($\phi^1,\phi^2$) is an
arbitrary complex spinor with $\bar{\phi}^i\phi_i=1$.  One can
define the SU(2) gauge field $a_a$ from
Eq.(\ref{Eq:so5_fundamental_spinor}) as $\bar{\psi}^\alpha d
\psi_\alpha=\bar{\phi}^\alpha(a_adx_a)_{\alpha\beta}\phi^\beta$,
where the dimensionless coordinate $x_a=X_a/R$ is used.  The field
strength $f_{ab}$ can be defined by $[D_a,D_b]$, where $D_a$ is
the covariant derivative.  Then, the Hamiltonian can be written as
$H=\frac{\hbar^2}{2MR^2}\sum_{a<b}\Lambda_{ab}^2$, where
$\Lambda_{ab}=-i(x_aD_b-x_bD_a)$.  Introducing
$L_{ab}=\Lambda_{ab}-if_{ab}$, the Hamiltonian can be expressed as
$H=\frac{\hbar^2}{2MR^2}(\sum_{a<b}L_{ab}^2-2I_i^2)$, where $I$
denotes the representation of the SU(2) gauge group. $L_{ab}$ can
be shown to satisfy the SO(5) algebra.  Therefore, the quantum
Hall states can be classified into the SO(5) representations
labelled by two integers $(p,q)$. Given $I$, $p$ can be related by
$p=2I+q$.  The spectrum of the generalized quantum Hall effect is
$E(2I+q,q)=\frac{\hbar^2}{2MR^2}(C(2I+q,q)-2I(I+1))$, where
$C(p,q)=p^2/2+q^2/2+2p+q$ is one of the Casimir operator of the
SO(5) group, and $q$ is the Landau level index.
\cite{Zhang2001Science}.

{\bf Larger symmtry in the lowest Landau level}  The lowest Landau
level (lll) is described by the SO(5) $(p,0)$ representation,
namely $q=0$.  The degeneracy is given by
$d(p,0)=\frac{1}{6}(p+1)(p+2)(p+3)$. The single-particle wave
function in the lll can be described only by the half of the
coordinates $\psi^\alpha$
\begin{eqnarray}
\sqrt{\frac{p!}{m_1!m_2!m_3!m_4!}}(\psi^1)^{m_1}(\psi^2)^{m_2}(\psi^3)^{m_3}(\psi^4)^{m_4}
\label{Eq:su4_coherent_state}
\end{eqnarray}
where $m_i$ are non-negative integers with $\sum_{i=1}^4 m_i =p$.
To have finite energy in the lll, $p$ has to be proportional to
$R^2$. The magnetic length $l_0$ can be defined as
$l_0=\text{lim}_{R\rightarrow \infty}R/\sqrt{p}$.  Furthermore, in
the large-$p$ limit, the degeneracy in the lll is proportional to
$p^3$, which is proportional to $R^6$.  It is because the SU(2)
gauge group introduces additional internal degrees of freedom
which is $S^2$. The total configuration space of lll counts from
the internal degrees of freedom $S^2$ and the orbital one $S^4$.
Locally, $S^4\times S^2$ is isomorphic to CP$^3$ which is the
six-dimensional complex projective space and the coordinates are
$(X_1,X_2,X_3,X_4,X_5)$ with $\sum_{i=1}^5 X_i=R^2$ for the
orbital and $(n_1,n_2,n_3)$ with $\sum_{i=1}^3 n_i = r^2$ for the
internal degrees of freedom.  From $\psi^\alpha$, $n_i$ is given
by $n_i/r =
\bar{\phi}^{\alpha}(\sigma_i)_{\alpha\beta}\phi^\beta$.
$\psi^\alpha$ actually describes a spinor on CP$^3$.  When the
number of particles $N=d(p,0)$, lll is fully filled. The many-body
wavefunction $\Psi$ is the Slater determinant of
Eq.(\ref{Eq:su4_coherent_state}) which is proportional to
\begin{eqnarray}
\Psi = \left| \begin{array}{ccccc} (\psi^1_1)^p &
(\psi^1_1)^{p-1}\psi^2_1 & . & . & (\psi^4_1)^p \\ (\psi^1_2)^p &
(\psi^1_2)^{p-1}\psi^2_2 & . & . & (\psi^4_2)^p \\ . & . & . & . &
.
\\ . & . & . & . & . \\ (\psi^1_N)^p & (\psi^1_N)^{p-1}\psi^2_N & . & . & (\psi^4_N)^p
\end{array} \right| \label{Eq:su4_IQHE}
\end{eqnarray}
where the subscripts are the particle indices.  For the fractional
states, ZH considered $\Psi^m$. The single-particle state becomes
$(mp,0)$\cite{Zhang2001Science}. While keeping the number of
particles $N=d(p,0)$ fixed, the filling factor $d(p,0)/d(mp,0)$
approaches to $1/m^3$ in the thermodynamic limit.

Besides SO(5) symmetry, the wavefunctions in the lll have larger
SU(4) symmetry, because SU(4) is the isometry group of CP$^3$. The
lll wavefunction can also be described by the SU(4)
representations which are denoted by three integers
$(n_1,n_2,n_3)$. Additionally, the lll wavefunction is described
by the SU(4) $(p,0,0)$ states with the degeneracy
$\frac{1}{6}(p+1)(p+2)(p+3)$ which is exactly the same as that of
SO(5) $(p,0)$ states. Furthermore, the SU(4) coherent states are
also given by Eq.(\ref{Eq:su4_coherent_state}).  In this case, the
single-particle state in the fractional case is described by the
SU(4) $(mp,0,0)$ state. Because we only consider the fractional
case in the lll, we do not care about the problem of the SO(5)
covariance.

Consider the following Hamiltonian
\begin{eqnarray}
H=\sum_{i<j}\sum_{q=1, \ \text{odd}}^{q\leq m\!-\!2}
\kappa_q~P_{ij}^{(2mp\!-\!2q,q,0)}. \label{Eq:pseudo-potential}
\end{eqnarray}
where $i$ and $j$ runs from 1 to $N$ and $\kappa_q$ are positive
constants.  $P_{ij}^{(2mp\!-\!2q,q,0)}$ indicate the projection
operator of the $(2mp\!-\!2q,q,0)$ states which describe the
two-fermion states when $q$ is odd. We will prove ZH fractional
quantum Hall state $\Psi^m$ is the zero-energy state of
Eq.(\ref{Eq:pseudo-potential}).

The two-fermion state is the antisymmetric channel of the
direct-product space of $(mp,0,0)\otimes(mp,0,0)$, which can be
decomposed as the direct-sum of the SU(4) invariant subspaces:
\begin{eqnarray}
(mp,0,0)\otimes(mp,0,0)|_a=\bigoplus_{q=1,\text{odd}}^{mp}(2mp-2q,q,0)
\label{Eq:su4_antisymmetric_CG}
\end{eqnarray}
where $a$ denotes the antisymmetric cannels.  For $m=1$, the
subspace in Eq.(\ref{Eq:su4_antisymmetric_CG}) with the highest
SU(3) weight is $(2p-2,1,0)$, because a general SU(4)
$(2p-2q,q,0)$ state can be decomposed as the direct sum of the
SU(3) states:
\begin{eqnarray}
&&(2p-2q,q,0)=(2p-q,0)+(2p-q-1,0)\nonumber\\&&+(2p-q-2,0)+..+(q,0)
\nonumber
\\&&+ (2p-q-1,1)+(2p-q-2,1)+..+(q-1,1) \nonumber \\&&+ ..
\nonumber
\\&&+(2p-2q,q)+(2p-2q-1,q)+..+(0,q) \label{Eq:su4_su3_decoposition}
\end{eqnarray}
In $\Psi^m$, the highest SU(3) weight is simply the
$m^{\text{th}}$ power of $(2p-1,0)$, namely $(2mp-m,0)$. The
two-fermion state in $\Psi^m$ contains only up to $(2mp-2m,m,0)$.
Therefore, it is the zero-energy state of the Hamiltonian in the
Eq.(\ref{Eq:pseudo-potential}).

{\bf Argument for the uniqueness} To show the non-degeneracy, let
us focus on the odd $p$ for convenience.  Particularly, we shall
use $p=3$ and $m=3$ as an example to illustrate our method.  In
this case, Eq.(\ref{Eq:pseudo-potential}) can be written as
$H=\kappa_1\sum_{i<j}P_{ij}^{(16,1,0)}$.  Let us call $\chi$ the
ground state wave function, namely $H\chi=0$.  Since we expect the
ground state wave function to be uniform, with the completeness of
the two-fermion states $\chi$ have to satisfy
\begin{eqnarray}
(P_{ij}^{(12,3,0)}+P_{ij}^{(8,5,0)}+P_{ij}^{(4,7,0)}+P_{ij}^{(0,9,0)})\chi=\chi
\ \ \forall (ij) \label{Eq:equation_for_p3}
\end{eqnarray}
To solve Eq.(\ref{Eq:equation_for_p3}), we expand $\chi$ in the
general form
\begin{eqnarray}
\sum_{\{\alpha_{jk}\}}\!\! C_3(\!\{\alpha_{jk}\!\})\!
\prod_{j=1}^{N}\prod_{k=1}^{9}\psi_j^{\alpha_{jk}}, \label{Eq:Phi}
\end{eqnarray}
where $N=20$ is the number of particles, and $\alpha_{jk}$ runs
from 1 to 4, and the summation goes over all configurations of
$\{\alpha_{jk}\}$, and finally
$\prod_{k=1}^{9}\psi_j^{\alpha_{jk}}$ is the $j^{\text{th}}$
single-particle state denoted by $(9,0,0)$, which is proportional
to Eq.(\ref{Eq:su4_coherent_state}).  From
Eq.(\ref{Eq:equation_for_p3}), $C_3$ satisfies the following
Schr\"odinger equation:
\begin{eqnarray}
&&C_3=\!\sum_{\{\beta_{ik},
\beta_{jk}\}}\!(A_3(\{\alpha_{ik}\},\{\alpha_{jk}\},\{\beta_{ik}\},\{\beta_{jk}\})\nonumber
\\&&+
A_5(\{\alpha_{ik}\},\{\alpha_{jk}\},\{\beta_{ik}\},\{\beta_{jk}\})\nonumber\\&&+A_7(\{\alpha_{ik}\},\{\alpha_{jk}\},\{\beta_{ik}\},\{\beta_{jk}\})\nonumber\\&&+A_9(\{\alpha_{ik}\},\{\alpha_{jk}\},\{\beta_{ik}\},\{\beta_{jk}\}))C_3
\ \ \forall (ij) \label{Eq:C2}
\end{eqnarray}
where the spinor indices of particles $i$ and $j$ are summed over,
and $A_q$ are the tensors denoting the Clebsch-Gordon (CG)
coefficients for the projection operators $P_{ij}^{(18-2q,q,0)}$.
In $(9,0,0)$ state, there are 9 symmetric indices.  To make the CG
coefficient from $(9,0,0)_i\times(9,0,0)_j$ to $(18-2q,q,0)_{ij}$,
we have to make $q$ antisymmetric pairs among the indices between
particles $i$ and $j$ and totally symmetrize the rest of the
indices.  For example, for a particular $(ij)$ $A_q$ can be given
as the following:
\begin{eqnarray}\nonumber
&&A_3(\{\alpha_{ik}\},\{\alpha_{jk}\},\{\beta_{ik}\},\{\beta_{jk}\})\nonumber\\&&=\!\frac{1}{N_3}(\delta_{\beta_{i1}}^{\alpha_{i1}}\delta_{\beta_{j1}}^{\alpha_{j1}}-\delta_{\beta_{j1}}^{\alpha_{i1}}\delta_{\beta_{i1}}^{\alpha_{j1}})(\delta_{\beta_{i4}}^{\alpha_{i4}}\delta_{\beta_{j4}}^{\alpha_{j4}}-\delta_{\beta_{j4}}^{\alpha_{i4}}\delta_{\beta_{i4}}^{\alpha_{j4}})\nonumber\\&&(\delta_{\beta_{i7}}^{\alpha_{i7}}\delta_{\beta_{j7}}^{\alpha_{j7}}-\delta_{\beta_{j7}}^{\alpha_{i7}}\delta_{\beta_{i7}}^{\alpha_{j7}})(\delta_{\beta_{i2}}^{\alpha_{i2}}\delta_{\beta_{i3}}^{\alpha_{i3}}\delta_{\beta_{i5}}^{\alpha_{i5}}\delta_{\beta_{i6}}^{\alpha_{i6}}\delta_{\beta_{i8}}^{\alpha_{i8}}\delta_{\beta_{i9}}^{\alpha_{i9}}\nonumber\\&&\delta_{\beta_{j2}}^{\alpha_{j2}}\delta_{\beta_{j3}}^{\alpha_{j3}}\delta_{\beta_{j5}}^{\alpha_{j5}}\delta_{\beta_{j6}}^{\alpha_{j6}}\delta_{\beta_{j8}}^{\alpha_{j8}}\delta_{\beta_{j9}}^{\alpha_{j9}}+\!\text{sym.})\nonumber
\\\nonumber &&
\\&&A_5(\{\alpha_{ik}\},\{\alpha_{jk}\},\{\beta_{ik}\},\{\beta_{jk}\})\nonumber\\&&=\!\frac{1}{N_5}(\delta_{\beta_{i1}}^{\alpha_{i1}}\delta_{\beta_{j1}}^{\alpha_{j1}}-\delta_{\beta_{j1}}^{\alpha_{i1}}\delta_{\beta_{i1}}^{\alpha_{j1}})(\delta_{\beta_{i2}}^{\alpha_{i2}}\delta_{\beta_{j2}}^{\alpha_{j2}}-\delta_{\beta_{j2}}^{\alpha_{i2}}\delta_{\beta_{i2}}^{\alpha_{j2}})\nonumber\\&&(\delta_{\beta_{i3}}^{\alpha_{i3}}\delta_{\beta_{j3}}^{\alpha_{j3}}-\delta_{\beta_{j3}}^{\alpha_{i3}}\delta_{\beta_{i3}}^{\alpha_{j3}})(\delta_{\beta_{i4}}^{\alpha_{i4}}\delta_{\beta_{j4}}^{\alpha_{j4}}-\delta_{\beta_{j4}}^{\alpha_{i4}}\delta_{\beta_{i4}}^{\alpha_{j4}})\nonumber\\&&(\delta_{\beta_{i7}}^{\alpha_{i7}}\delta_{\beta_{j7}}^{\alpha_{j7}}-\delta_{\beta_{j7}}^{\alpha_{i7}}\delta_{\beta_{i7}}^{\alpha_{j7}})(\delta_{\beta_{i5}}^{\alpha_{i5}}\delta_{\beta_{i6}}^{\alpha_{i6}}\delta_{\beta_{i8}}^{\alpha_{i8}}\delta_{\beta_{i9}}^{\alpha_{i9}}\delta_{\beta_{j5}}^{\alpha_{j5}}\delta_{\beta_{j6}}^{\alpha_{j6}}\nonumber\\&&\delta_{\beta_{j8}}^{\alpha_{j8}}\delta_{\beta_{j9}}^{\alpha_{j9}}+\!\text{sym.})\nonumber
\\&&\nonumber\\&&A_7(\{\alpha_{ik}\},\{\alpha_{jk}\},\{\beta_{ik}\},\{\beta_{jk}\})\nonumber\\&&=\!\frac{1}{N_7}(\delta_{\beta_{i1}}^{\alpha_{i1}}\delta_{\beta_{j1}}^{\alpha_{j1}}-\delta_{\beta_{j1}}^{\alpha_{i1}}\delta_{\beta_{i1}}^{\alpha_{j1}})(\delta_{\beta_{i2}}^{\alpha_{i2}}\delta_{\beta_{j2}}^{\alpha_{j2}}-\delta_{\beta_{j2}}^{\alpha_{i2}}\delta_{\beta_{i2}}^{\alpha_{j2}})\nonumber\\&&(\delta_{\beta_{i3}}^{\alpha_{i3}}\delta_{\beta_{j3}}^{\alpha_{j3}}-\delta_{\beta_{j3}}^{\alpha_{i3}}\delta_{\beta_{i3}}^{\alpha_{j3}})(\delta_{\beta_{i4}}^{\alpha_{i4}}\delta_{\beta_{j4}}^{\alpha_{j4}}-\delta_{\beta_{j4}}^{\alpha_{i4}}\delta_{\beta_{i4}}^{\alpha_{j4}})\nonumber\\&&(\delta_{\beta_{i5}}^{\alpha_{i5}}\delta_{\beta_{j5}}^{\alpha_{j5}}-\delta_{\beta_{j5}}^{\alpha_{i5}}\delta_{\beta_{i5}}^{\alpha_{j5}})(\delta_{\beta_{i6}}^{\alpha_{i6}}\delta_{\beta_{j6}}^{\alpha_{j6}}-\delta_{\beta_{j6}}^{\alpha_{i6}}\delta_{\beta_{i6}}^{\alpha_{j6}})\nonumber\\&&(\delta_{\beta_{i7}}^{\alpha_{i7}}\delta_{\beta_{j7}}^{\alpha_{j7}}-\delta_{\beta_{j7}}^{\alpha_{i7}}\delta_{\beta_{i7}}^{\alpha_{j7}})(\delta_{\beta_{i8}}^{\alpha_{i8}}\delta_{\beta_{i9}}^{\alpha_{i9}}\delta_{\beta_{j8}}^{\alpha_{j8}}\delta_{\beta_{j9}}^{\alpha_{j9}}+\!\text{sym.})\nonumber
\\&&\nonumber\\&&A_9(\{\alpha_{ik}\},\{\alpha_{jk}\},\{\beta_{ik}\},\{\beta_{jk}\})\nonumber\\&&=\!\frac{1}{N_9}(\delta_{\beta_{i1}}^{\alpha_{i1}}\delta_{\beta_{j1}}^{\alpha_{j1}}-\delta_{\beta_{j1}}^{\alpha_{i1}}\delta_{\beta_{i1}}^{\alpha_{j1}})(\delta_{\beta_{i2}}^{\alpha_{i2}}\delta_{\beta_{j2}}^{\alpha_{j2}}-\delta_{\beta_{j2}}^{\alpha_{i2}}\delta_{\beta_{i2}}^{\alpha_{j2}})\nonumber\\&&(\delta_{\beta_{i3}}^{\alpha_{i3}}\delta_{\beta_{j3}}^{\alpha_{j3}}-\delta_{\beta_{j3}}^{\alpha_{i3}}\delta_{\beta_{i3}}^{\alpha_{j3}})(\delta_{\beta_{i4}}^{\alpha_{i4}}\delta_{\beta_{j4}}^{\alpha_{j4}}-\delta_{\beta_{j4}}^{\alpha_{i4}}\delta_{\beta_{i4}}^{\alpha_{j4}})\nonumber\\&&(\delta_{\beta_{i5}}^{\alpha_{i5}}\delta_{\beta_{j5}}^{\alpha_{j5}}-\delta_{\beta_{j5}}^{\alpha_{i5}}\delta_{\beta_{i5}}^{\alpha_{j5}})(\delta_{\beta_{i6}}^{\alpha_{i6}}\delta_{\beta_{j6}}^{\alpha_{j6}}-\delta_{\beta_{j6}}^{\alpha_{i6}}\delta_{\beta_{i6}}^{\alpha_{j6}})\nonumber\\&&(\delta_{\beta_{i7}}^{\alpha_{i7}}\delta_{\beta_{j7}}^{\alpha_{j7}}-\delta_{\beta_{j7}}^{\alpha_{i7}}\delta_{\beta_{i7}}^{\alpha_{j7}})(\delta_{\beta_{i8}}^{\alpha_{i8}}\delta_{\beta_{j8}}^{\alpha_{j8}}-\delta_{\beta_{i8}}^{\alpha_{j8}}\delta_{\beta_{j8}}^{\alpha_{i8}})\nonumber\\&&(\delta_{\beta_{i9}}^{\alpha_{i9}}\delta_{\beta_{j9}}^{\alpha_{j9}}-\delta_{\beta_{j9}}^{\alpha_{i9}}\delta_{\beta_{i9}}^{\alpha_{j9}})
\label{Eq:zero-energy}
\end{eqnarray}
where $\text{sym.}$ means to totally symmetrize the lower indices
of the Kronecker delta function $\delta^\alpha_\beta$, and $N_q$
are the normalization constants.  For any pair $(ij)$. the indices
to be set antisymmetric in $A_q$ are arbitrary because
$\{\alpha_{jk}\}$ are totally symmetric for each particle $j$.
Moreover, from Eq.(\ref{Eq:equation_for_p3}) every $A_q$ makes
\emph{at least} 3 and odd number of antisymmetric pairs.
Therefore, we find the symmetry that from
Eq.(\ref{Eq:zero-energy}) $C_3$ becomes $-C_3$ by exchanging whole
group of indices $(\alpha_{i1}\alpha_{i2}\alpha_{i3})$ and
$(\alpha_{j1}\alpha_{j2}\alpha_{j3})$, or by exchanging
$(\alpha_{i4}\alpha_{i5}\alpha_{i6})$ and
$(\alpha_{j4}\alpha_{j5}\alpha_{j6})$, or by exchanging
$(\alpha_{i7}\alpha_{i8}\alpha_{i9})$ and
$(\alpha_{j7}\alpha_{j8}\alpha_{j9})$.  If this symmetry is true
for any pair $(ij)$, the argument of the non-degeneracy can be
given as the following: first, we fix the indices
$(\alpha_{i4}\alpha_{i5}..\alpha_{i9})$ for all particles, namely
$i=1,..,N$.  The symmetry that $C_3$ picks up a minus sign when
exchanging $(\alpha_{i1}\alpha_{i2}\alpha_{i3})$ and
$(\alpha_{j1}\alpha_{j2}\alpha_{j3})$ for any pair $(ij)$ demands
the total antisymmetry of the wave function for the first three
indices.  On the other hand, the dimension spanned by the first
three indices is nothing but $\frac{1}{6}\cdot 4\cdot 5\cdot 6$,
which is equal to the number of particles $N=20$.  Then, there is
a unique wave function $\Psi$ given by Eq.(\ref{Eq:su4_IQHE}) for
the first three indices of every particle, namely
$C_3\sim\epsilon_{(\alpha_{11}\alpha_{12}\alpha_{13})(\alpha_{21}\alpha_{22}\alpha_{23})..(\alpha_{N
1}\alpha_{N2}\alpha_{N3})}$, where $\epsilon$ is the
totally-antisymmetric tensor with respect to exchanging indices of
whole $\alpha_{i1}\alpha_{i2}\alpha_{i3}$ and
$\alpha_{j1}\alpha_{j2}\alpha_{j3}$.  Similarly, applying this
argument to the second three and the third three indices of every
particle, we obtain $\Psi^3$ as the unique ground state wave
function for $\chi$.

Now, suppose that there exist another ground state wave function
that does not satisfy the symmetry mentioned above for any pair
$(ij)$.  Let us think of the easiest case.  Namely, the symmetry
is satisfied for any pair $(ij)$ except for the pair $(i'j')$ that
under the exchange of $(\alpha_{i'1}\alpha_{i'2}\alpha_{i'3})$ and
$(\alpha_{j'1}\alpha_{j'2}\alpha_{j'3})$,
\begin{eqnarray}
C'_3=\sum[-A_3+A_5-A_7-A_9]C_3\label{Eq:unique}
\end{eqnarray}
where we omit the summation variables.  From Eq.(\ref{Eq:unique})
and Eq.(\ref{Eq:C2}), we can define another variable $C"=C_3-C'_3$
for all $(ij)$.  It is now clear that the solution given by $C"$
does not have the $(8,5,0)$ state for the pair $(i'j')$.  However,
$C"$ satisfies the symmetry mentioned above.  According to the
argument above, it has to be $\Psi^3$. Then, the contradiction
occurs, because one can easily show that $\Psi^3$ contains
$(8,5,0)$ state for any pair $(ij)$. To reconcile the
contradiction, $\Psi^3$ has to be the unique solution. This
argument can be generalized to more complicate cases or general
$p$ cases.

{\bf The excitation with fractional charges}  The natural
generalization of Laughlin's quasi-particle/hole operators are
given as the following\cite{Haldane1983PRL}.
\begin{eqnarray}
&&
B^\dag_N(\Phi_{\alpha})=\prod_{i=1}^{N}(\Phi_{\alpha}\mathcal{R}_{\alpha\beta}\psi^i_\beta)\label{Eq:su4_cp3_hole-like_excitation}
\\ && B_N(\Phi_{\alpha})=\prod_{i=1}^{N}(\Phi^*_\alpha\mathcal{R}_{\alpha\beta}\frac{\partial}{\partial
\psi^i_\beta}) \label{Eq:su4_cp3_particle-like_excitation}
\end{eqnarray}
where $\Phi_{\alpha}$ is a four-component complex spinor with
$\bar{\Phi}^{\alpha}\Phi_{\alpha}=1$ denoting the position that
the excitation is created in CP$^3$.  $\mathcal{R}_{\alpha\beta}$
is the charge conjugate matrix which takes the following form
\begin{eqnarray}
\mathcal{R}_{\alpha\beta}= \left( \begin{array}{cc} -i\sigma_2 & 0
\\ 0 & -i\sigma_2 \end{array} \right)
\end{eqnarray}
$B^\dag_N(\Phi_{\alpha})\Psi^m_N$ ($B_N(\Phi_{\alpha})\Psi^m_N$)
describes a hole-like (particle-like) excitation because the size
of the system has been enlarged (reduced) by
$+(-)\frac{1}{2}m^2p^2$, where the single-particle state is
described by the $(mp+(-)1,0,0)$ state. Because $p\!\sim\! R^2$,
these excitations are the four-dimensional objects, namely
quasi-4-branes.  The extended excitations are not very new to
condensed matter physicists. For example, in superfluid, a vortex
excitation is a point-like particle in 2 spatial dimensions and a
one-dimensional string in 3 spatial dimensions.  In general, in
$D$ spatial dimensions, a vortex is a ($D$-2)-dimensional extended
objects. In our case, CP$^3$ is 6-dimensional.  The quantum
quasi-4-brane may be regarded as a vortex excitation in the
generalized fractional quantum Hall fluid \cite{Zhang1989PRL,
Reed1989PRL, bernevig-2002-300}.

We can apply Haldane's argument of the fractional charge in our
system\cite{Haldane1983PRL}.  In the thermodynamical limit, the
number of particle $N \sim\frac{1}{6}p^3$, so
$p\sim\sqrt[3]{6}N^{\frac{1}{3}}$. Define
$p(N,m)=\sqrt[3]{6}mN^{\frac{1}{3}}$ for the fractional case such
that the single-particle state is in the $(p(N,m),0,0)$ state.  By
changing the field strength, a state with $N^{\text{ex}}_p$
particle-like and $N^{\text{ex}}_h$ hole-like quasi-4-branes has
$p=p(N,m)+(N^{\text{ex}}_h-N^{\text{ex}}_p)$. On the other hand,
if we fix the field strength and excite the systems by removing
(injecting) particles, we have to remove (inject)
$\frac{1}{2}m^2p^2\sim\frac{(\sqrt[3]{6})^2}{2}m^2N^{\frac{2}{3}}$
particles to make the quasi-4-brane.  Then, we obtain
$p(N\pm\frac{(\sqrt[3]{6})^2}{2}m^2N^{\frac{2}{3}},m)=\sqrt[3]{6}m(N\pm\frac{(\sqrt[3]{6})^2}{2}m^2N^{\frac{2}{3}})^{\frac{1}{3}}\simeq
\sqrt[3]{6}mN^{\frac{1}{3}}\pm m^3$.  In other words, one 4-brane
with charge $+(-)1$ is attached by $m^3$ flux.  Then, the
elementary excitation by changing unit flux carries a fractional
charge $q=+(-)1/m^3$.

Finally, let us discuss about the density correlation function in
the fractional case defined by $\rho_m(x,x')=\frac{1}{(N-2)!}\int
dx_3\cdot\cdot dx_N|\Psi^m_N|^2$.  For $m=1$
\cite{Zhang2001Science}, it is given by
\begin{eqnarray}
\rho_1(x,x')=1-|\bar{\psi}_{\alpha}(x)\psi_\alpha(x')|^{2p}.\label{Eq:density}
\end{eqnarray}
  When one take $x'$ to be the north pole of both the orbital and
the internal space and let $x$ approach to $x'$, then
$\rho_1(x,x')\sim 1-e^{-\frac{1}{4l_0^2}(X^2_{\mu}+N^2_{\alpha})}$
provided that $R=r$ where $X^2_{\mu}=R^2\sum_{\mu=1}^4x^2_{\mu}$
and $N^2_{\alpha}=R^2(n^2_1+n^2_2)$\cite{Zhang2001Science}.  We
calculate $\rho_3(x,x')$ for $m=3$ and $p=1$:
\begin{eqnarray}
\rho_3(x,x')=(1-|\bar{\psi}_{\alpha}(x)\psi_\alpha(x')|^{2})^3+O(x_{\mu}^8,n_{\alpha}^8)
\label{Eq:density_2}
\end{eqnarray}
when $x$ approaches $x'$.  As two particles are close enough, the
higher order vanishes faster than the leading order term.
Comparing Eq.(\ref{Eq:density}) and Eq.(\ref{Eq:density_2}), the
density correlation function for general $p$ and $m$ should have
the following form
\begin{eqnarray}
\rho_m(x,x')&\sim&
(1-|\bar{\psi}_{\alpha}(x)\psi_\alpha(x')|^{2p})^m \nonumber
\\&\sim& (1-e^{-\frac{1}{4l_0^2}(X^2_{\mu}+N^2_{\alpha})})^m.
\label{Eq:su4_density_correlation}
\end{eqnarray}
Eq.(\ref{Eq:su4_density_correlation}) states that in the filling
factor $\nu=1/m^3$ case the density correlation function vanishes
as \emph{$m$-th order root} when two particles approach to each
other.  This behavior suggests that the generalized fractional
quantum Hall state is an incompressible liquid.

To summarize, similar to the 2D fractional quantum Hall effect,
the 4D FQHS has to be stabilized by a two-body repulsion
interaction. We also prove the non-degeneracy of the 4D FQHS.  We
found the correspondent elementary excitations of  Laughlin's
quasi-hole (quasi-particle) in CP$^3$, which can be considered as
vortices in CP$^3$. They are extended objects carrying fractional
charges $+(-)1/m^3$. We also discuss the density correlation
function in the fractional case, which indicates that the FQHS is
an incompressible liquid.

There are several open issues regarding to the FQHS.  In
Ref.\cite{bernevig-2002-300}, using the effective field-theoretic
approach the FQHS can be shown to support 2-dimensional
excitations which obey fractional statistics.  The correspondent
states in our pseudo-potential approach need to be clarified.
Also, the edge mode of the FQHS is always an intriguing issue for
the future exploration.

We are grateful for the stimulating discussions with Darwin Chang,
Dung-Hai Lee, and Naoto Nagaosa.  This work is supported by the
ERATO-SSS project.
%\bibliography{4dfqhe}

\begin{thebibliography}{12}
\expandafter\ifx\csname
natexlab\endcsname\relax\def\natexlab#1{#1}\fi
\expandafter\ifx\csname bibnamefont\endcsname\relax
  \def\bibnamefont#1{#1}\fi
\expandafter\ifx\csname bibfnamefont\endcsname\relax
  \def\bibfnamefont#1{#1}\fi
\expandafter\ifx\csname citenamefont\endcsname\relax
  \def\citenamefont#1{#1}\fi
\expandafter\ifx\csname url\endcsname\relax
  \def\url#1{\texttt{#1}}\fi
\expandafter\ifx\csname
urlprefix\endcsname\relax\def\urlprefix{URL }\fi
\providecommand{\bibinfo}[2]{#2}
\providecommand{\eprint}[2][]{\url{#2}}

\bibitem[{\citenamefont{Zhang and Hu}(2001)}]{Zhang2001Science}
\bibinfo{author}{\bibfnamefont{S.-C.} \bibnamefont{Zhang}} \bibnamefont{and}
  \bibinfo{author}{\bibfnamefont{J.}~\bibnamefont{Hu}},
  \bibinfo{journal}{Science} \textbf{\bibinfo{volume}{294}},
  \bibinfo{pages}{823} (\bibinfo{year}{2001}).

\bibitem[{\citenamefont{Murakami et~al.}(2003)\citenamefont{Murakami, Nagaosa,
  and Zhang}}]{Murakami2003Science}
\bibinfo{author}{\bibfnamefont{S.}~\bibnamefont{Murakami}},
  \bibinfo{author}{\bibfnamefont{N.}~\bibnamefont{Nagaosa}}, \bibnamefont{and}
  \bibinfo{author}{\bibfnamefont{S.-C.} \bibnamefont{Zhang}},
  \bibinfo{journal}{Science} \textbf{\bibinfo{volume}{301}},
  \bibinfo{pages}{1348} (\bibinfo{year}{2003}).

\bibitem[{\citenamefont{Bernevig et~al.}(2002)\citenamefont{Bernevig, Chern,
  Hu, Toumbas, and Zhang}}]{bernevig-2002-300}
\bibinfo{author}{\bibfnamefont{B.~A.} \bibnamefont{Bernevig}},
  \bibinfo{author}{\bibfnamefont{C.-H.} \bibnamefont{Chern}},
  \bibinfo{author}{\bibfnamefont{J.-P.} \bibnamefont{Hu}},
  \bibinfo{author}{\bibfnamefont{N.}~\bibnamefont{Toumbas}}, \bibnamefont{and}
  \bibinfo{author}{\bibfnamefont{S.-C.} \bibnamefont{Zhang}},
  \bibinfo{journal}{ANNALS PHYS.} \textbf{\bibinfo{volume}{300}},
  \bibinfo{pages}{185} (\bibinfo{year}{2002}).

\bibitem[{\citenamefont{Bernevig et~al.}(2003)\citenamefont{Bernevig, Hu,
  Toumbas, and Zhang}}]{bernevig-2003-91}
\bibinfo{author}{\bibfnamefont{B.~A.} \bibnamefont{Bernevig}},
  \bibinfo{author}{\bibfnamefont{J.~P.} \bibnamefont{Hu}},
  \bibinfo{author}{\bibfnamefont{N.}~\bibnamefont{Toumbas}}, \bibnamefont{and}
  \bibinfo{author}{\bibfnamefont{S.~C.} \bibnamefont{Zhang}},
  \bibinfo{journal}{Physical Review Letters} \textbf{\bibinfo{volume}{91}},
  \bibinfo{pages}{236803} (\bibinfo{year}{2003}).

\bibitem[{\citenamefont{Fabinger}(2002)}]{Fabinger2002JHEP}
\bibinfo{author}{\bibfnamefont{M.}~\bibnamefont{Fabinger}},
  \bibinfo{journal}{JHEP} \textbf{\bibinfo{volume}{0205}}, \bibinfo{pages}{037}
  (\bibinfo{year}{2002}).

\bibitem[{\citenamefont{Elvang and Polchinski}(2002)}]{Henriette2002}
\bibinfo{author}{\bibfnamefont{H.}~\bibnamefont{Elvang}} \bibnamefont{and}
  \bibinfo{author}{\bibfnamefont{J.}~\bibnamefont{Polchinski}},
  \bibinfo{journal}{hep-th/0209104}  (\bibinfo{year}{2002}).

\bibitem[{\citenamefont{Hu and Zhang}(2002)}]{Hu2002PRB}
\bibinfo{author}{\bibfnamefont{J.}~\bibnamefont{Hu}} \bibnamefont{and}
  \bibinfo{author}{\bibfnamefont{S.-C.} \bibnamefont{Zhang}},
  \bibinfo{journal}{Phys. Rev. B} \textbf{\bibinfo{volume}{66}},
  \bibinfo{pages}{125301} (\bibinfo{year}{2002}).

\bibitem[{\citenamefont{Kimura}(2002)}]{Kimura2002NPB}
\bibinfo{author}{\bibfnamefont{Y.}~\bibnamefont{Kimura}},
  \bibinfo{journal}{Nucl. Phys. B} \textbf{\bibinfo{volume}{637}},
  \bibinfo{pages}{177} (\bibinfo{year}{2002}).

\bibitem[{\citenamefont{Laughlin}(1983)}]{Laughlin1983PRL}
\bibinfo{author}{\bibfnamefont{R.}~\bibnamefont{Laughlin}},
  \bibinfo{journal}{Phys. Rev. Lett.} \textbf{\bibinfo{volume}{50}},
  \bibinfo{pages}{1395} (\bibinfo{year}{1983}).

\bibitem[{\citenamefont{Haldane}(1983)}]{Haldane1983PRL}
\bibinfo{author}{\bibfnamefont{F.}~\bibnamefont{Haldane}},
  \bibinfo{journal}{Phys. Rev. Lett.} \textbf{\bibinfo{volume}{51}},
  \bibinfo{pages}{605} (\bibinfo{year}{1983}).

\bibitem[{\citenamefont{Zhang et~al.}(1989)\citenamefont{Zhang, Hansson, and
  Kivelson}}]{Zhang1989PRL}
\bibinfo{author}{\bibfnamefont{S.-C.} \bibnamefont{Zhang}},
  \bibinfo{author}{\bibfnamefont{T.}~\bibnamefont{Hansson}}, \bibnamefont{and}
  \bibinfo{author}{\bibfnamefont{S.}~\bibnamefont{Kivelson}},
  \bibinfo{journal}{Physical Review Letters} \textbf{\bibinfo{volume}{62}},
  \bibinfo{pages}{82} (\bibinfo{year}{1989}).

\bibitem[{\citenamefont{Reed}(1989)}]{Reed1989PRL}
\bibinfo{author}{\bibfnamefont{N.}~\bibnamefont{Reed}},
  \bibinfo{journal}{Physical Review Letters} \textbf{\bibinfo{volume}{62}},
  \bibinfo{pages}{88} (\bibinfo{year}{1989}).

\end{thebibliography}

\end{document}